\documentclass[11pt,a4paper,twocolumn]{article}    
\usepackage{graphicx}
\usepackage{amsmath}
\usepackage{amssymb}
\usepackage{xcolor}
\usepackage{url}
\usepackage{hyperref}
\usepackage{natbib}      
\usepackage{journals_aas}      

\def\fracd#1#2{\frac{\displaystyle #1}{\displaystyle #2}}

\begin{document}

\title{\vskip -2em \underline{\small Astronomy Reports, 2024, Vol. 68, No. 10, pp. 1022-1028. doi 10.1134/S1063772924700884} \\[2em]
  Should We Expect Further Acceleration of the Earth's Rotation\\ in the Coming Years?}
\author{Z. M. Malkin\\ Pulkovo Observatory, Russian Academy of Sciences, St. Petersburg, 196140 Russia\\ e-mail: malkin@gaoran.ru\\[1em]
Received April 16, 2024; revised July 8, 2024; accepted August 15, 2024}
\date{~}

\twocolumn[
\begin{@twocolumnfalse}
\maketitle
\begin{abstract}
Recently, it has been suggested in the literature that the difference between universal and coordinated
time UT1--UTC could reach a large positive value in the coming years \citep{Agnew2024}.
This would make it necessary
to introduce a negative leap second into UTC for the first time in history, which in turn will cause serious
problems in time keeping and synchronization systems around the world. Based on the latest Earth's rotation
and universal time data published by the international Earth rotation and reference systems service (IERS)
and their prediction, in this paper, we have shown that the acceleration trend observed over the past four years
is likely to return to slowing down soon. Therefore, fears about the possible need to introduce a negative leap
second into the UTC time scale in the next few years in the light of recent observational data have seen
unfounded.\\
\vskip 0ex\noindent
\textbf{Keywords}: Earth's rotation, Earth's rotation speed, time scales,
universal time\\
\vskip 0ex\noindent
\textbf{DOI}: 10.1134/S1063772924700884
\end{abstract}
\end{@twocolumnfalse}
\vspace{2em}
]


\section{INTRODUCTION}

Almost all time scales used by humanity are somehow
connected with observations of the Earth's rotation.
The main time scale directly related to the
Earth's rotation is the universal time (UT1), which is
determined by the angular rotation of the Earth
around its axis of rotation relative to the celestial reference
system \citep{McCarthy1991IEEEP,Zharov2006book,IERSConv2010}.
Thus, UT1 is an astronomical time
scale. However, the Earth's rotation is a very complex
and, generally speaking, non-stationary geophysical
process, which causes the unevenness of the UT1 time
scale. This makes the UT1 unsuitable for most everyday
practical applications.

Another time scale widely used for storing and disseminating
time is international atomic time (TAI),
which is based on a combination (averaging) of the
time scales of hundreds of atomic clocks operating in
dozens of laboratories located around the world \citep{Guinot2005}.
TAI is a very uniform and even time scale, but it is also
not always convenient for universal use due to the
divergence from UT1 astronomical time, which
increases with time. The TAI--UT1 difference, which
was $\sim$1.4~s in 1961, now exceeds~37~s.

To overcome these problems and provide a more
suitable time scale that, on the one hand, would be
close to the angle of rotation of the Earth and, on the
other hand, would be as homogeneous as possible, a
new time scale, coordinated universal time (UTC),
was introduced in 1961 \citep{Nelson2001,Panfilo2019}.
UTC is an atomic time
that is the same rate as TAI but differs from it by
an integer number of seconds (after January~1, 1972;
before this date, the difference between TAI and UTC
was calculated using a more complex procedure).
According to the latest international agreement, the
absolute value of the difference between UT1 and
UTC must not exceed 0.9~s. This is monitored by the
International Earth Rotation and Reference Systems
Service (IERS), which is responsible for introducing
the leap second, usually at the end of June or the end
of December, when it is necessary to compensate for
the accumulated difference between TAI and UTC.
The end of March and September could also be
reserve dates for introducing the leap second, but they
have never been used yet. Thus, UTC is a stepped time
scale (see Fig.~\ref{fig:c04_ut1-utc} below).
UTC is currently the primary
time scale for civil use in most countries of the world.

The difference between the astronomical universal
time scale UT1 and the atomic time scale TAI, usually
denoted as TAI--UT1, increased monotonically from
1961 (when the UTC scale was introduced) until the
early 2020s (excluding small decadal and seasonal
variations), after which an anomalous acceleration of
the Earth's rotation began to be observed, which was
reflected in a decreasing trend in TAI--UT1 in recent
years (see Fig.\ref{fig:c04_tai-ut1} below).

During the period of steady slowing of the Earth's
rotation, all leap seconds introduced so far into the
UTC time scale were ``positive'', and they were
inserted between the moments of 23$^h$53$^m$59$^s$ of correction
dates and 0$^h$0$^m$00$^s$ of the next day. Over the years,
users of the UT1 and UTC time scales have adapted to
this procedure.

If the trend of the Earth's rotation acceleration
continues for a long time, it may be necessary to introduce
a ``negative'' second into UTC, which could lead
to serious disruptions in time storage and synchronization
systems. This scenario was recently discussed in \citet{Agnew2024}.
In this paper, we attempt to determine based on
recent observational data to what extent these concerns
are substantiated.


\section{ANALYSIS OF DATA ON UNIVERSAL TIME UT1}
\label{sect:ut1_analysis}

This study is based on the analysis of the UT1 series
calculated by IERS. The IERS structure consists of
several components, including several centers for data
analysis and calculation of various consolidated decisions
(Product Centers). One of them, the Earth Orientation
Center, located at the Paris Observatory
(OPA) calculates the IERS Earth orientation parameter
(EOP) series based on a combination of various data
obtained by several space geodesy techniques.

\begin{figure*}
\centering
\includegraphics[clip,width=0.9\textwidth]{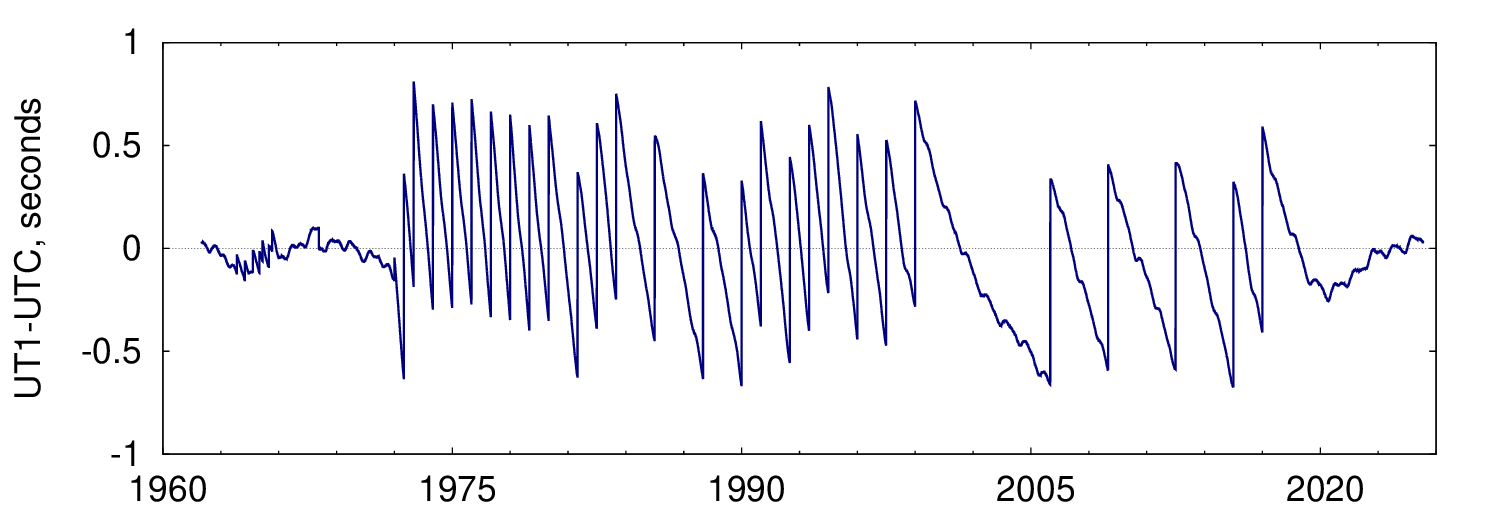}
\caption{IERS C04 UT1--UTC series.}
\label{fig:c04_ut1-utc}
\end{figure*}

The main series of IERS EOP can probably be considered the C04 EOP series,
which is most widely used in current scientific research and practical
applications \citep{Bizouard2019}.
Series C04 is updated daily and contains
daily values of eight EOP (coordinates of the Earth's
pole and their rate of change, coordinates of the celestial
pole, UT1--UTC, and the length of the day) at 0h
of the date starting from January~1, 1962. The last
epoch of this series falls 30 days back relative to the
current date (the date of publication of the C04 series
on the IERS website). The IERS C04 UT1--UTC
series is shown in Fig.~\ref{fig:c04_ut1-utc}.

Figure~\ref{fig:c04_ut1-utc} shows how the atomic time scale UTC
is adjusted to the astronomical time scale UT1, keeping
the difference between the UT1 and UTC scales
within the limits of $\pm$0.9~s. Between January~1, 1961
and December~31, 1971, 13~minor adjustments were
made to the UTC scale, including changes in its rate.
Therefore, this section of the UTC scale is not of great
interest for scientific analysis and further consideration
in this paper. Since 1972, the UTC time scale has
been adjusted only by introducing whole leap seconds.
A total of 28 extra seconds have been introduced since
the beginning of 1972, the last of which was introduced
on December~31, 2017.

The normal behavior of the UT1--UTC time scale
difference until the early 2020s was its decrease over
time, not counting small deviations from monotonicity
due to seasonal and decadal variations in the
Earth's rotation speed. Therefore, until now, an additional
second has always been introduced when
UT1--UTC approaches the lower limit of the permissible
difference between UT1 and UTC, i.e., to
with some time reserve since this procedure is always
done in advance and announced several months
before the actual adjustment of the UTC scale.
This time reserve is necessary for users of the universal
and coordinated time scales to prepare equipment and
software in advance for the jump to UTC. Therefore,
in practice, the introduction of an additional second
occurs when UT1--UTC approaches --0.7~s (Fig.~\ref{fig:c04_ut1-utc}).

\begin{figure*}[p]
\centering
\includegraphics[clip,width=0.9\textwidth]{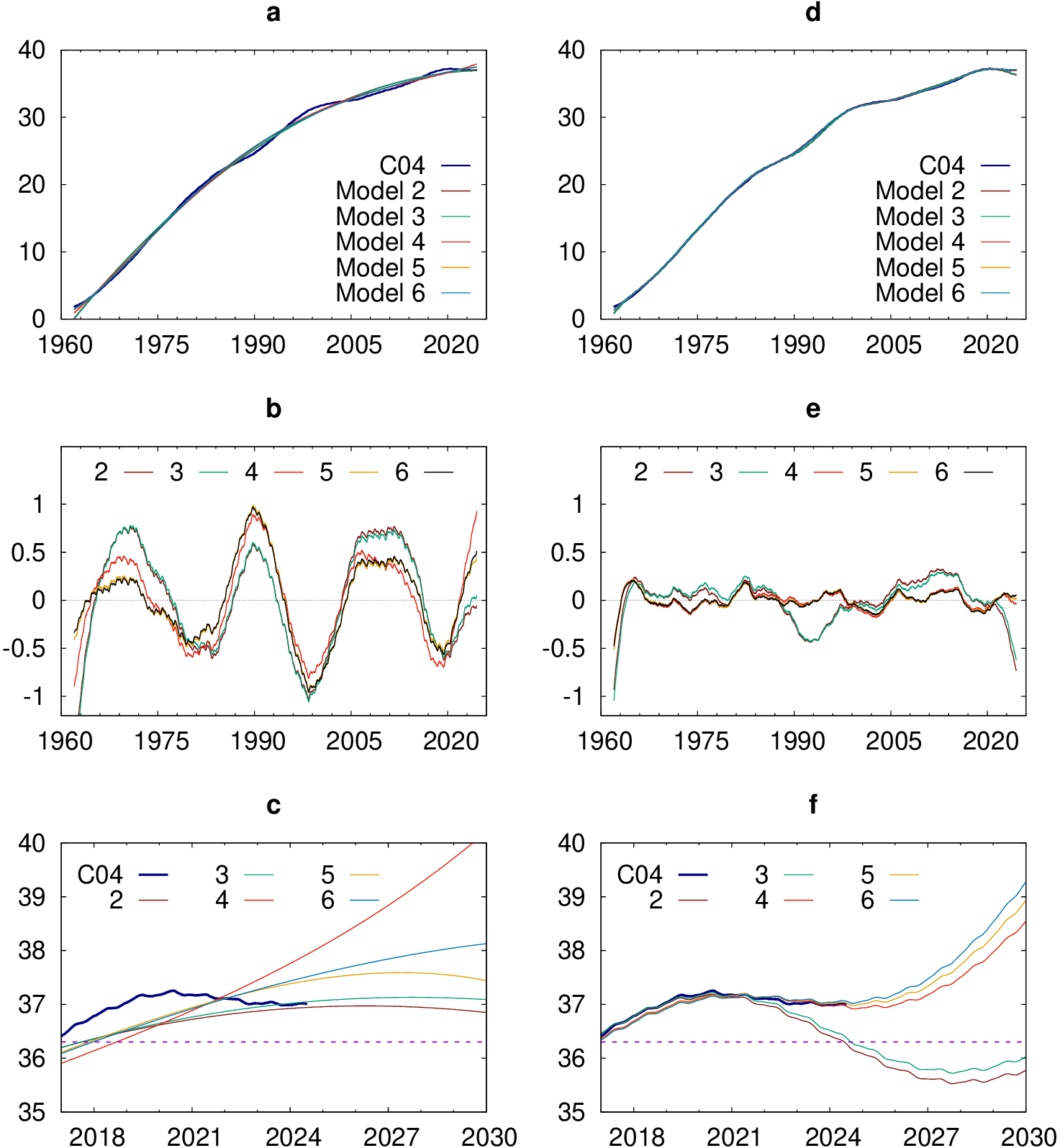}
\caption{Results of approximation of TAI--UT1 changes (in seconds) over the period 1962--2024. The left column corresponds to
the approximation by a polynomial of degree 2 to 6 (the order of the polynomial is indicated in the captions to the graphs), the
right column corresponds to the approximating model consisting of a polynomial of degree 2 to 6 and four harmonics. In each
column, the graphs show: at the top--a comparison of the IERS C04 series with the approximating model, in the center--the
differences between the model and observations, at the bottom--the last seven years of data from the graphs in the top row extrapolated
to 2030. In the lower graphs, the dotted line indicates the approximate threshold value of TAI--UT1 (36.3~s) for making a
decision on introducing a negative leap second.}
\label{fig:c04_tai-ut1}
\end{figure*}

The recent disruption of this rhythm due to a rather
abrupt transition to a significant acceleration of the
Earth's rotation, starting around 2020, has led to the
suggestion (see, e.g., \citet{Agnew2024}) about the probable
need to introduce a negative leap second into the UTC time
scale at the end of the current decade, which would be
the first case in the history of the UTC time scale.
Below, this question is examined in more detail in
order to estimate how realistic this scenario is in light
of recent astronomical observations of the Earth's
rotation.

On the top panel of Fig.~\ref{fig:c04_tai-ut1}, the TAI--UT1 series
is presented, which is obtained from the C04 UT1--UTC series
using the expression:
\begin{equation}
\begin{array}{l}
\mathrm{TAI-UT1} = \\ \quad \mathrm{(TAI-UTC)-(UT1-UTC)} \,,
\end{array}
\end{equation}
where TAI--UTC is the correction to the UTC scale that is also distributed
by the IERS EOP calculation centers. Currently, TAI--UTC is 37~s.

The C04 TAI--UT1 series was then fitted with a polynomial
model to determine the global trend in universal
time over the last 50+ years of UTC. The calculations
were carried out with polynomials from the second to
the sixth degree. The results of this simulation are
shown in the left panel of Fig.\ref{fig:c04_tai-ut1},
which presents a comparison of the model with the C04 series
(top panel), the ``model minus C04'' offsets (middle
panel), and the results of extrapolating the model to
2030 (bottom panel).

The need to introduce a negative leap second into
UTC may arise when TAI--UT1 decreases to a value
less than approximately 36.3~s or, equivalently, when
UT1--UTC increases to a value greater than approximately
0.7~s. The results obtained with the polynomial
model showed that all model variants do not predict
reaching this threshold before the early 2030s. At the
same time, it should be noted that the polynomial
model is not the best forecasting method over a forecast
horizon of several years because it assumes the
preservation of the general long-term trend and does
not take into account the latest observational data,
which can critically affect the short-term (compared
to the length of the entire C04 series) forecast.

In Fig.~\ref{fig:c04_tai-ut1}a, decadal variations are clearly visible
against the background of a smooth change in TAI--UT1.
Therefore, an attempt was made to take them
into account in a refined model, to which four harmonic
components with periods of 18.613, 12, 1, and
0.5~years were added, which are discussed in more
detail below. The results of applying the polynomial--harmonic
model are shown in the right part of Fig.~\ref{fig:c04_tai-ut1}.
It is clear that this model describes the real variations
of universal time much better. In this case, close
results were obtained for polynomials of the second
and third degrees on the one hand and for polynomials
of degrees from 4 to 6 on the other hand. The second
group of models showed significantly better approximation
accuracy, especially at the end of the series,
where the approximating functions with polynomials
of degree~2 and~3 showed significantly larger differences
from the real data, which can critically affect the
accuracy of the forecast, which is especially important
for this study.

The extrapolated data presented in Fig.~\ref{fig:c04_tai-ut1}f
indeed show an unsatisfactory result for the second- and
third-degree polynomial model variants: they diverge
sharply from the actual data, predicting the need to
introduce a negative leap second in UTC as early as
mid-2024, which is clearly not the case. Therefore, for
the final calculations, the option with the smallest
polynomial order in the second group, the fourth, was
selected. Thus, the final model used here is as follows:
\begin{equation}
\begin{array}{l}
(\mathrm{TAI-UT1})_{mod} = \sum\limits_{i=0}^4 a_i^p t^i\\
\quad + \, \sum\limits_{i=1}^4 \left(a_i^s\sin\fracd{2\pi t}{P_i}
   + a_i^c\cos\fracd{2\pi t}{P_i}\right) \,,
\end{array}
\label{eq:model}
\end{equation}
where $t=t_{C04}-t_0$, $t_{C04}$ is the epoch of the IERS C04
series in years, $t_0$ is the average epoch of the series
(with this choice of the initial epoch, a minimum of
coefficient errors is ensured), and $P_i$ is periods of harmonics
in years. The coefficients of the formula
approximating the TAI--UT1 series over the entire
interval from January~1, 1962 to June~5, 2024 are given
in Table~\ref{tab:approximation}.

\begin{table}
\caption{Coefficients of the approximation formula for TAI--UT1 (in s).}
\label{tab:approximation}
\begin{tabular}{cr@{$\cdot$}lr@{$\cdot$}l}
\hline
 Coefficient & \multicolumn{2}{c}{Value} & \multicolumn{2}{c}{Error} \\
\hline\\[-2.5ex]
$a_0^p$ & $ 2.7525$ & $10^{1} $ & \quad 1.1096 & $10^{-3}$ \\[0.5ex]
$a_1^p$ & $ 5.8075$ & $10^{-1}$ & \quad 8.1162 & $10^{-5}$ \\[0.5ex]
$a_2^p$ & $-1.1503$ & $10^{-2}$ & \quad 7.5677 & $10^{-6}$ \\[0.5ex]
$a_3^p$ & $ 9.1062$ & $10^{-6}$ & \quad 1.2736 & $10^{-7}$ \\[0.5ex]
$a_4^p$ & $ 2.8159$ & $10^{-6}$ & \quad 9.1339 & $10^{-9}$ \\[0.5ex]
$a_1^s$ & $ 5.7398$ & $10^{-1}$ & \quad 8.5065 & $10^{-4}$ \\[0.5ex]
$a_1^c$ & $-2.9036$ & $10^{-1}$ & \quad 9.4552 & $10^{-4}$ \\[0.5ex]
$a_2^s$ & $ 1.8905$ & $10^{-1}$ & \quad 8.5578 & $10^{-4}$ \\[0.5ex]
$a_2^c$ & $-6.9516$ & $10^{-2}$ & \quad 8.6960 & $10^{-4}$ \\[0.5ex]
$a_3^s$ & $ 1.0162$ & $10^{-2}$ & \quad 8.2469 & $10^{-4}$ \\[0.5ex]
$a_3^c$ & $ 2.1670$ & $10^{-2}$ & \quad 8.2520 & $10^{-4}$ \\[0.5ex]
$a_4^s$ & $ 7.4285$ & $10^{-3}$ & \quad 8.2493 & $10^{-4}$ \\[0.5ex]
$a_4^c$ & $-5.3899$ & $10^{-3}$ & \quad 8.2468 & $10^{-4}$ \\[0.5ex]
\hline
\end{tabular}
\end{table}

Among the harmonic components of the model,
the first harmonic stands out with a period of
18.613~years and an amplitude of 0.64~s associated with
the tide corresponding to the period of precession of
the lunar orbit. It has been previously discussed in the
literature \citep{Ray2014,LeMouel2019,Zotov2020}.
The annual and semi-annual periodicities
in the Earth's rotation rate have also been
well known for a long time. These two harmonics are
added to the model for completeness; they have virtually
no effect on the results of the study due to the
smallness of their amplitudes: 24 and 9~ms, respectively,
but may be of independent interest for studying
seasonal variations in universal time.

A harmonic with a period of 12~years and an amplitude
of 0.2~s was found empirically in this study. Inclusion
of such a harmonic in the model ensured a reduction
in residual differences compared to the option of
using a harmonic with a period of 11~years associated
with the Schwabe cycle of solar activity. The presence
of an 11-year periodicity in the Earth's rotation rate
has also been noted in previous studies \citep{LeMouel2019}.
Apparently, the 12-year harmonic has accumulated some
additional variations of universal time. On the other
hand, the average duration of the solar cycle over the
last 50~years (i.e., over the period of considered dates)
is slightly greater than 11 years\footnote{\url{https://www.sidc.be/SILSO/cyclesmm}}.
In any case, a detailed study of this issue is beyond the scope of this paper.

Harmonics with the same periods of 18.6 and
12.0~years and with amplitudes of 0.63~s and 0.20~s,
respectively, which is close to the results of the present
study, were also found by \citet{Tissen2021}
from processing a 100-year series of observations.

In the changes in universal time, the general tendency
towards a gradual decrease in the rate of growth
of TAI--UT1 is primarily distinguished. However, it is
premature to consider a transition to a stage of significant
decrease in TAI--UT1, at least until the beginning of the 2030s,
given the complex and poorly predictable behavior of the Earth's
rotation speed in the past \citep{Stephenson2016,Morrison2021}.
Also, the available observational data show alternation
of periods of relative acceleration and deceleration
of the Earth's rotation with a main period of
$\sim$18.6~years. The amplitude of these decadal oscillations
is not very stable and changes approximately in
the range from 0.5 to 1~s. An increase in the TAI--UT1
difference corresponds to a deceleration of the Earth's
rotation while its decrease corresponds to an acceleration
of the Earth's rotation. Accordingly, an increase
in the UT1--UTC difference, on the contrary, corresponds
to an acceleration of the Earth's rotation while
its decrease corresponds to a slowdown of the Earth's
rotation.

Periods of relative acceleration of the Earth's rotation
against the background of the general trend are
observed in 1985--1990, in 2000--2005, and, finally, in
the current period after 2020. Thus, the time intervals,
during which the relative acceleration of the Earth's
rotation is observed, last about five years. During
these periods, the introduction of a leap second into
UTC is required less frequently, as can be seen in
Fig.~\ref{fig:c04_ut1-utc}.
It is interesting to note that periods of decreasing
Earth's rotation speed last longer than periods of
its increase, which in itself is interesting and deserves a
separate study.

From the data presented in Fig.~\ref{fig:c04_ut1-utc},
it can be seen that the rate of increase in the difference between
the UT1 and UTC time scales has significantly decreased
in the last one or two years compared to the early
2020s. Therefore, it is interesting to estimate the possible
behavior of UT1 in the near future, using the
methods of forecasting the UT1--UTC time scales
that are well developed in specialized services.

Besides the long-term final IERS EOP series calculated
at OPA, another IERS center, the Rapid Service/Prediction Center\footnote{\url{https://maia.usno.navy.mil/}},
located at the United States
Naval Observatory (USNO), calculates the official
operational IERS EOP data with a one-year forecast
\citep{McCarthy1991,Luzum2001,Stamatakos2020jsrs,IERSAR2019}.

\begin{figure*}[t]
\centering
\includegraphics[clip,width=0.9\textwidth]{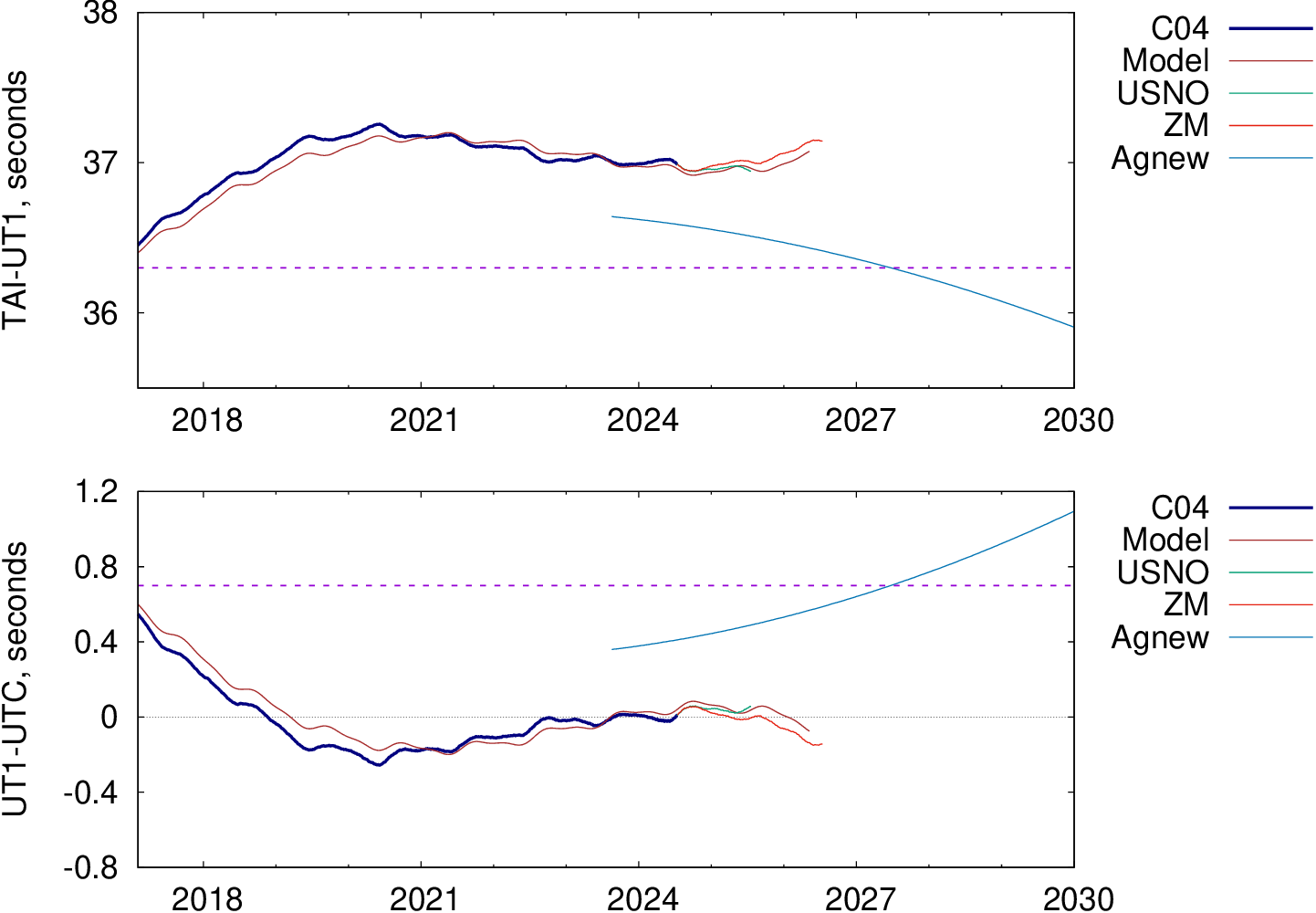}
\caption{Last seven years of the IERS C04 universal time series as TAI--UT1
(upper panel) and UT1--UTC (lower panel) with the USNO annual forecast,
the author's biennial forecast (ZM), and the forecast from \citet{Agnew2024}.
The brown line corresponds to the approximating
model described by formula (2).
The dotted lines indicate the approximate threshold values of TAI--UT1 (36.3~s)
and UT1--UTC (0.7 s) for making a decision to introduce a negative leap second.}
\label{fig:tai-ut1_with_predictions}
\end{figure*}

In Fig.~\ref{fig:tai-ut1_with_predictions},
the last seven years of the IERS C04
world time series are shown with the USNO annual
forecast and with the two-year forecast calculated by
the author using the method described in \citet{Malkin1996}.
The figure also shows the result of extrapolation of the
approximating function given by formula~(\ref{eq:model}), with the
coefficients given in the Table 1, for the next two years.
The longer-term forecast does not look reasonable.
The agreement between the forecast based on the
extrapolation of the fitting function (brown line) and
the forecasts calculated by other methods in USNO
and by the author (ZM) is remarkably good.

Forecast graph from \citep{Agnew2024} in Fig.~\ref{fig:tai-ut1_with_predictions}
is based on the numerical files included with the online version of this
paper. It is taken into account that both the paper and
the files actually provide data not for TAI, as is indicated,
but for TT = TAI + 32.184~s (CA Agnew, private
communication). It is clearly visible that this forecast
shows a significant shift along the vertical axis of the
beginning of the forecast relative to the end of the
IERS C04 series by $\sim$0.3~s that was also observed in
\citet[Fig. 2d]{Agnew2024}.
It is believed that this shift may be
caused by incompleteness of the geophysical model
used to describe the actual rotation of the Earth.

Although the results of the universal time predictions
shown in Fig.~\ref{fig:tai-ut1_with_predictions} vary somewhat,
they all predict that the acceleration of the Earth's rotation will
cease in the near future and do not anticipate a resumption
of the acceleration of the Earth's rotation until the 2030s.
This contradicts the conclusion made in the study \citet{Agnew2024},
which, on the contrary, assumes an increase in the acceleration
of the Earth's rotation speed after 2023--2024.


\section{Conclusions}
\label{sect:conclusions}

In this study, the behaviour of UT1 in recent
decades has been studied using IERS Earth's rotation
data and their forecast. As a result of this study, it has
been shown that the anomalous acceleration of the
Earth's rotation that has been observed since the early
2020s has slowed down significantly recently. The current
forecast for universal time has suggested that the
period of acceleration of the Earth's rotation must end
within the next two years and will most likely be
replaced by a slowdown, which is more typical of the
behavior of the Earth's rotation speed in recent
decades, as well as in more distant retrospect,
although against the background of decadal fluctuations
\cite{Stephenson2016,Morrison2021}.

Thus, the assumption made in \cite{Agnew2024} about the
expected increasing speed of the Earth's rotation in
the coming years and, as a consequence, about the
possible need for the first time since the beginning of
the UTC time scale to introduce a negative leap second
into it in the second half of the 2020s, apparently
does not have sufficient grounds. The reason for such
a significant discrepancy between the results of the
present study and the results \cite{Agnew2024}, apparently,
is the incompleteness of the geophysical model accepted
in \cite{Agnew2024} that does not reflect with sufficient accuracy
the real features of the Earth's rotation, as well as a different
approach to predicting universal time.

Several recent universal time forecasts presented in
Fig.~\ref{fig:tai-ut1_with_predictions} do not predict
a significant decrease in the TAI--UT1 difference
in the second half of the 2020s, but
rather show an emerging trend towards a transition to
a stage of its growth. Thus, the results obtained in this
study are a significant refinement of the result \cite{Agnew2024}
in light of recent observational data.

From the consideration of the entire 60-year series
of universal time used in this study, it is evident that
the results of astronomical observations show a general
tendency towards a slowing down of the Earth's rotation
rate. This is reflected in the increase in the difference
between the TAI--UT1 time scales. However, the
growth rate of this difference decreases over time and
the TAI--UT1 value even decreased slightly after 2020
(Fig.~\ref{fig:c04_tai-ut1}).
However, it is clearly premature to consider a
long-term trend towards a further reduction of this difference.
The latest universal time forecasts made by
the author and the USNO indicate that if the trend
and decadal changes in the Earth's rotation rate
remain stable in the coming years, it is highly likely
that the upper limit of the permissible UT1--UTC difference
will not be reached in the coming years. We are
currently on the descending branch of the 18-year
cycle (for TAI--UT1, see Figs.~\ref{fig:c04_tai-ut1}
and \ref{fig:tai-ut1_with_predictions}),
which must soon be replaced by the ascending one.

In this regard, it has been also interesting to note
the results obtained in \cite{Zotov2020,Zotov2023}.
The authors have studied
decadal variations in the length of day (LOD) over
the time interval 1830--2020 and identified a harmonic
component with a period of $\sim$60~years and a
significant amplitude of $\sim$2~ms (which, according to
the authors' assumption, may be a superposition of
90- and 20--40-year oscillations). According to these
results, this (quasi-)60-year wave is the main contributor
to the deviation of the observed LOD changes
from the linear trend determined by the Earth's secular
tidal deceleration. Currently, this decadal wave in
LOD is at a minimum and a new period of increasing
LOD and, therefore, slowing down the Earth's rotation,
must begin soon. This is consistent with the
results of the present study and provides an additional
explanation for the observed variations in universal
time in recent years and their current forecast.

Based on all of the above, it can be assumed that
the next approach of the UT1--UTC difference to the
upper limit of $\sim$0.7~s (or the TAI--UT1 difference to
the lower limit of ~36.3~s) will occur no earlier than the
beginning of the 2030s. At the same time, the plan to
revise the strategy for maintaining the UTC time scale,
which has been actively discussed in recent years,
(Resolution 4 of the 27th CGPM ``On the use
and future development of
UTC''\footnote{\url{https://www.bipm.org/en/cgpm-2022/resolution-4/}},
\citep{ITU2023} if is implemented,
it may make this problem irrelevant by that time.


\section*{ACKNOWLEDGMENTS}

The author thanks three reviewers for very helpful comments
and suggestions on the initial version of the article.
The SAO/NASA Astrophysics Data System abstract database\footnote{\url{https://ui.adsabs.harvard.edu/}}
(ADS) was used in preparing the paper.
The figures were prepared using the program \texttt{gnuplot}\footnote{\url{http://www.gnuplot.info/}}.

\section*{FUNDING}

This work was supported by ongoing institutional funding.
No additional grants to carry out or direct this particular research were obtained.

\section*{CONFLICT OF INTEREST}

The author of this work declares that he has no conflicts of interest.

\section*{OPEN ACCESS}

This article is licensed under a Creative Commons Attribution
4.0 International License, which permits use, sharing,
adaptation, distribution and reproduction in any medium or
format, as long as you give appropriate credit to the original
author(s) and the source, provide a link to the Creative Commons
license, and indicate if changes were made. The images
or other third party material in this article are included in the
article's Creative Commons license, unless indicated otherwise
in a credit line to the material. If material is not included
in the article's Creative Commons license and your intended
use is not permitted by statutory regulation or exceeds the
permitted use, you will need to obtain permission directly
from the copyright holder. To view a copy of this license, visit
\url{http://creativecommons.org/licenses/by/4.0/}


\bibliography{ut1_now.bib}
\bibliographystyle{joge}

\end{document}